\documentclass[twocolumn]{aastex631}

\usepackage{xspace}

\usepackage{graphicx}
\usepackage{dcolumn}
\usepackage{bm}
\usepackage[nointegrals]{wasysym}
\usepackage{verbatim}
\usepackage{color}
\usepackage{amsmath}
\usepackage[utf8]{inputenc}
\usepackage[shortlabels]{enumitem}
\usepackage{wrapfig}
\usepackage{scalerel}
\usepackage[caption=false]{subfig}
\usepackage{multirow} 

\usepackage{graphicx}	
\usepackage{amsmath}	
\usepackage{booktabs}

\newcommand*\planck{\emph{Planck}\xspace}
\newcommand*\lcdm{$\Lambda$CDM\xspace}

\newcommand*\lite{\texttt{lite}\xspace}
\newcommand*\cut{\texttt{cut}\xspace}
\newcommand*\npipe{\texttt{NPIPE}\xspace}
\newcommand*\plik{\texttt{Plik}\xspace}
\newcommand*\camspec{\texttt{CamSpec}\xspace}
\newcommand*\pliklegacy{\texttt{Plik-Legacy}\xspace}
\newcommand*\camspeclegacy{\texttt{CamSpec-Legacy}\xspace}
\newcommand*\camspecnpipe{\texttt{CamSpec-NPIPE}\xspace}
\newcommand*\mflike{\texttt{MFLike}\xspace}
\newcommand*\cobaya{\texttt{Cobaya}\xspace}
\newcommand*\pact{\texttt{P-ACT}\xspace}
\newcommand*\camb{\texttt{camb}\xspace}

\newcommand{\onreview}[1]{#1} 

\begin{document}

\title[\emph{The choice of \emph{Planck} in cosmological analyses}]{The choice of \emph{Planck} CMB likelihood in cosmological analyses}


\author{Hidde T. Jense}\thanks{E-mail: jenseh@cardiff.ac.uk}\affiliation{School of Physics and Astronomy, Cardiff University, The Parade, CF24 3AA Cardiff, Wales, UK}

\author{Marc Vi\~na}\affiliation{School of Physics and Astronomy, Cardiff University, The Parade, CF24 3AA Cardiff, Wales, UK}

\author{Erminia Calabrese}\affiliation{School of Physics and Astronomy, Cardiff University, The Parade, CF24 3AA Cardiff, Wales, UK}

\author{J.~Colin~Hill}\affiliation{Department of Physics, Columbia University, New York, NY 10027, USA}

\date{\today}



\begin{abstract}
We compare cosmological parameters from different \planck sky maps and likelihood pipelines, assessing robustness of cosmological results with respect to the choice of the latest \planck maps-likelihood combination. We show that, for the \planck multipole range retained in combination with ground-based observations, different products give very similar cosmological solutions; small remaining differences are reduced by the addition of other CMB datasets to \planck. In particular, constraints on extended cosmological models benefit from the addition of small-scale power from ground-based experiments and are completely insensitive to the choice of \planck maps and likelihood. For this work we derive and release a nuisance-marginalized dataset and \camspecnpipe-\lite likelihood for the \planck NPIPE data injected into the CamSpec likelihood --- which are usually used to obtain the reference \planck PR4 cosmology. Using the extracted CMB spectra we show that the additional constraining power for cosmology is coming from polarization at all scales and from temperature at multipoles above 1500 when going from PR3 to PR4. We also show that full marginalization over the CamSpec foreground nuisance parameters can impact parameter inference and model selections when truncating some scales; our new likelihood enables correct combinations with other CMB datasets. \\ 
\end{abstract}




\section{Introduction}

The cosmic microwave background (CMB) has been a primary source of constraints on cosmological models for decades. Until recently, the most stringent limits on the parameters of the standard cosmological model, \lcdm, and its extensions were set using observations of CMB temperature and polarization anisotropies from the \planck\ satellite mission~\citep{Planck_overview_2018,Planck_spec_like_2018,Planck_params_2018,rosenberg:2022,hillipop2024}. Constraints on cosmological parameters similar or better than those obtained with \planck\ are now becoming possible with data from the ground-based Atacama Cosmology Telescope (ACT;~\citealp{dr6maps,dr6ps,dr6ext}) and South Pole Telescope (SPT;~\citealp{Camphuis:2025uvg}), exploiting information encoded in the CMB at intermediate and small scales --- multipoles between ${400\lesssim \ell \lesssim 6500}$. However, \planck\ remains the dominant dataset in tracing the behaviour of the very large scales, at ${2<\ell<1600}$. This range of observations from \planck is currently used in combination with ACT and SPT to derive new leading cosmological results and will continue to supplement large-scale data to future ground-based experiments until a new CMB satellite becomes operational.

The \planck\ products available to the community for cosmological exploitation are based on two sets of sky maps. The `Legacy' maps~\citep{Planck_overview_2018} are processed through the \texttt{Plik} likelihood described in~\cite{Planck_spec_like_2018} (PL20 hereafter) with cosmology results reported in~\cite{Planck_params_2018} (PC20 hereafter). The Legacy maps are also analyzed in~\cite{CamSpec_legacy} (EG21 hereafter) with the alternative \camspec likelihood pipeline. The `NPIPE' maps~\citep{Planck_NPIPE} were generated after introductions of several data-processing improvements in the handling of large-scale systematics subsequent to the final \emph{Planck} legacy release. The reduction in systematics allowed a small increase in the sky fraction used for cosmology, and revised cosmological limits from these maps have been presented in~\cite{rosenberg:2022} (R22 hereafter) with the \camspec likelihood pipeline and in~\cite{hillipop2024} with the \texttt{HiLLiPoP} likelihood. 

\begin{table*}[htb]
    \centering
    \begin{tabular}{l|c c l l}
    \toprule
    \textbf{\planck Products} & \textbf{Likelihood} & \textbf{Maps} & \textbf{Reference} & \textbf{Known as}\\
    \midrule
    \pliklegacy & \plik & Legacy & \citet{Planck_overview_2018,Planck_spec_like_2018,Planck_params_2018} & PR3 reference cosmology \\
    \camspeclegacy & \camspec & Legacy & \citet{CamSpec_legacy} & \\
    \camspecnpipe & \camspec & \npipe & \citet{rosenberg:2022} & PR4 reference cosmology\\
    \texttt{HiLLiPoP-NPIPE} & \texttt{HiLLiPoP} & \npipe & \citet{hillipop2024} &\\
    \bottomrule
    \end{tabular}
    \caption{Summary of \planck products available for cosmological exploitation, making explicit the choice of sky maps and likelihood pipelines. PR3 cosmology is usually drawn from the final legacy release of \planck, using the Legacy maps processed with the \texttt{Plik} likelihood; PR4 cosmology is usually drawn from the analysis of NPIPE maps processed with the \texttt{CamSpec} likelihood.\\}
    \label{tab:likelihood-data-overview}
\end{table*}

The two sets of maps define the \planck release names, with Legacy corresponding to the \planck data Release 3, PR3, and NPIPE maps identified as PR4. However, the reference cosmology for each data release relies also on the choice of the likelihood. PR3 cosmology is then usually referring to the Legacy maps processed with \texttt{Plik}, and PR4 cosmology usually points to NPIPE maps processed with \texttt{CamSpec}. A summary of these products is presented in~\autoref{tab:likelihood-data-overview}.

In this paper, we compare and discuss differences between the reference PR3 and PR4 cosmological results, using the Legacy maps in input to the \plik likelihood code (hereafter labeled \pliklegacy) and the NPIPE maps in input to the \camspec likelihood (hereafter labeled \camspecnpipe). While both sets of maps have been processed with the \camspec likelihood, the details of the likelihood method applied to Legacy and NPIPE maps in \camspec differ (see EG21 and R22), and a \plik-\texttt{NPIPE} likelihood is not available.
When using the full dataset the cosmology retrieved from the \pliklegacy and \camspecnpipe likelihoods are similar but show order of $1\sigma$ movements in cosmological parameters, as discussed in R22. Some of these shifts are due to different choices in data characterization made between \plik and \camspec and present even when using the same Legacy maps --- see discussion in PL20, PC20 and EG21. Additionally, \camspecnpipe provides more stringent constraints on some cosmological parameters, due to the increased sky fraction retained in the NPIPE maps (R22) corresponding to $\sim 10\%$ more data.
Here we extract and use the CMB signal-to-noise in temperature and polarization as metric to show where the additional constraining power for cosmology in \camspecnpipe is localized. We also add to previous comparisons, focusing in particular on the restricted \planck multipole range used in combined analyses with ACT and SPT --- to e.g., form the `\pact' combination of~\cite{dr6ps} and `CMB-SPA' combination of~\cite{Camphuis:2025uvg} --- to assess the robustness of new, leading constraints on cosmological models with respect to the choice of \planck likelihoods and maps.

To perform these studies we work with foreground and nuisance-marginalized likelihoods, deriving and releasing a new \camspecnpipe-\lite likelihood. We show that attention to marginalization over nuisances is required when truncating the \camspecnpipe likelihood and necessary for likelihood combinations. 

The paper is organized as follows. In \autoref{sec:likelihoods} we summarize the main aspects of the \plik and \camspec likelihoods and present a new, foreground-marginalized \camspecnpipe likelihood; \autoref{sec:cosmology} compares and discusses the constraints on cosmological parameters from various likelihoods; finally, in \autoref{sec:p-act-cosmology}, we derive and discuss cosmological constraints from the combination of the various \planck likelihoods with small-scale measurements from ACT DR6. We conclude in~\autoref{sec:conclusions}.

\section{Planck likelihoods} \label{sec:likelihoods}
\subsection{Likelihood data and model}

Both \plik and \camspec likelihoods take in input versions of cross-frequency angular power spectra from the \planck's High Frequency Instrument (HFI), together with a covariance matrix capturing the instrumental noise properties of the multi-frequency maps. The two likelihoods differ in amount and format of data, as well as in the detailed modelling of the spectra --- as explained in~\cite{Planck_spec_like_2015}, PL20, EG21, R22. In this work we do not alter any pre-processing and modelling done before or at likelihood level. To compare and discuss cosmological results between different likelihood packages, we rely on the underlying CMB-only power in the data. To extract this we simply need to add an additional processing step of the multi-frequency data and operate on the foregrounds and systematics components accounted for in the likelihood. In what follows we then summarize only the main aspects relevant for this work and refer the reader to PL20, EG21 and R22 for the full details.

All likelihood evaluations in this work are performed within the Simons Observatory\footnote{\url{https://simonsobservatory.org/}} \mflike framework\footnote{\url{https://github.com/simonsobs/LAT_MFLike}}, which provides consistent handling of input spectra, covariances, and marginalization procedures. A version of \pliklegacy in this framework was presented in~\cite{Li23}\footnote{Available at \url{https://github.com/simonsobs/LAT_MFLike/tree/mflike-plik}}; the equivalent for \camspecnpipe is built in this work\footnote{Available at \url{https://github.com/MarcVB93/mflike-camspec}}.

\subsubsection{\pliklegacy}
The \texttt{Plik} likelihood uses a total of 16 auto- and cross-spectra in TT, TE, and EE constructed from T and E Legacy half-mission maps at frequencies 100, 143, and 217~GHz. The spectra only cover the range of multipoles over which the likelihood of the CMB power spectra can be well-approximated as Gaussian and span multipoles $30<\ell<2508$ in temperature, and $30<\ell<1996$ in polarization --- the upper and lower multipoles are different across spectra and the number of cross-frequency spectra is different between temperature and TE/EE (PL20). The spectra are binned for a total of 613 datapoints in the full temperature plus polarization dataset. 

Before computing spectra, the maps are masked to clean for Galactic dust and bright point sources in temperature, and only Galactic dust in polarization. This leaves the likelihood to include Galactic and extragalactic foreground emission. The model of foregrounds in temperature scales across frequencies templates for the thermal and kinetic Sunyaev--Zel'dovich effects (tSZ and kSZ), the cosmic infrared background (CIB), tSZxCIB correlations, and solves for unresolved point sources as a different varying shot noise term in each cross-spectrum.
Thermal dust emission from the Galaxy is estimated by cross-correlating each map individually with the 545 GHz map, building a template for each cross-spectrum that takes into account the correct sky fraction and frequency-dependent point source mask that differs per map. The resulting template is then only left to scale via an amplitude, which is constrained with a Gaussian prior during parameter inference.
In polarization, extragalactic emission is limited to a point source contribution, which is negligible for the sensitivity of \planck. Polarized dust emission is estimated similarly to the thermal dust emission, except that the cross-correlation is done with the 353 GHz channel (which is the highest \planck frequency channel that measures polarization), and the template used is a simple power law with a fixed spectral index and an amplitude that is free to vary only for the TE cross-spectrum. The EE spectra are not very sensitive to the value of the dust amplitude, and these are left fixed to the value found from the cross-correlation with the 353 GHz channel.

Residual systematics such as beam leakage, sub-pixel effects, and correlated noise corrections are also accounted for at likelihood level with fixed frequency-dependent templates.

The \plik likelihood function comes down to a Gaussian likelihood
\begin{equation}
    -2 \ln \mathcal{L}(\theta) = \left( \mathcal{D}_b^{\rm th}(\theta) - \mathcal{D}_b^{\rm d} \right) \mathbf{\Sigma}^{-1} \left( \mathcal{D}_b^{\rm th}(\theta) - \mathcal{D}_b^{\rm d} \right)^T ,
\end{equation}
where for each cross-spectrum XY covering TT, TE, EE, $\mathcal{D}_b^{\rm d}$ is the binned data vector, $\mathbf{\Sigma}$ the covariance matrix, and $\mathcal{D}_b^{\rm th}(\theta)$ the binned theory model defined by some parameters $\theta$, built from the components
\begin{eqnarray}
    \mathcal{D}_b^{{\rm th}, XY}(\theta) & =& \sum_{\ell} \mathbf{w}_b^{\ell} c^X_{\nu_1} c^Y_{\nu_2} \Big( \mathcal{D}_\ell^{{\rm CMB}, XY}(\theta_1) \nonumber\\
    & & + \mathcal{D}_{\ell,\nu_1,\nu_2}^{{\rm fg}, XY}(\theta_2) + \mathcal{D}_\ell^{{\rm sys}, XY} \Big)
\end{eqnarray}
where $\mathbf{w}_b^\ell$ is the binning matrix used by \plik, $\mathcal{D}_\ell^{\rm CMB}$ is the CMB theory model, $\mathcal{D}_{\ell,\nu_1,\nu_2}^{\rm fg}$ is the foreground model at cross-frequency $\nu_1 \times \nu_2$, and $\mathcal{D}_\ell^{\rm sys}$ is the residual systematic correction. The final data vector contains binned data, where each bin $b$ is specific to a cross spectrum $XY$, frequencies $\nu_1, \nu_2$, and $\ell$ range. Here, we divide the full parameter vector $\theta$ between the cosmological parameters $\theta_1$ and a set of secondary emission parameters $\theta_2$ accounting for foregrounds. The components of the theory vector are corrected by factors $c^{X}_{\nu_1}$, $c^Y_{\nu_2}$, which contain all relevant calibration and polarization efficiency factors for each cross-spectrum --- some of them left as free parameters in the likelihoods and others fixed to measured quantities. \pliklegacy defines the calibration relative to the 143 GHz temperature map, leaving five free parameters\onreview{: two} for the calibration of the 100 and 217 GHz maps relative to this, and three for the polarization efficiencies of the polarization maps. The 143 GHz map is calibrated off the dipole measurement, and the uncertainty in this measurement is captured in a total calibration parameter $A_{\rm Planck}$, which linearly scales all spectra. The full posterior mode also contains contributions from parameter priors, most of which are wide, uniform priors with little information for parameters that are well-constrained by the data. Gaussian priors are imposed on the Galactic dust emission amplitudes from the measurement at 545 GHz, and the overall calibration factor $A_{\rm Planck} = (100 \pm 0.25) \%$ propagating the uncertainty in the dipole calibration.

The official \pliklegacy product is also released in a compressed form (PL20), known as `\plik-\lite' and labeled throughout here as \pliklegacy-\lite, which marginalizes over foreground and calibration parameters and systematics templates to produce a reduced data vector that retains the CMB-only cosmological constraining power while improving computational efficiency. 


\subsubsection{\camspecnpipe}
The \camspecnpipe\ likelihood 
is constructed from five unbinned angular power spectra: three TT cross-spectra ($143\times143$, $143\times217$, and $217\times217$\,GHz), and coadded TE and EE spectra built from the \npipe maps. The $143\times143$\,GHz TT spectrum, as well as the TE and EE spectra, span the multipole range $30 < \ell < 2000$, while the $143\times217$ and $217\times217$\,GHz TT spectra cover $500 < \ell < 2500$. The polarization spectra are formed by inverse-noise-variance weighting of all cross-spectra between 100, 143, and 217\,GHz; and only cross-\onreview{split} spectra are used in the construction of the likelihood, avoiding the noise bias associated with auto-spectra. \onreview{This means any likelihood spectrum is built up only of A$\times$B spectra, where A and B are the different detector set splits chosen by \npipe.}

One of the main difference with \pliklegacy\ --- beyond the use of different sets of maps --- is that significant foreground removal is done at map level and not modelled in the likelihood. 
This pre-cleaning includes both masking and the use of additional frequency channels to suppress foreground contamination. For example, a relatively conservative Galactic mask is applied to the NPIPE maps, especially at low Galactic latitudes, to reduce contamination from diffuse foregrounds. This is sufficient to render post-cleaning dust residuals negligible, even at the most contaminated frequency used in this likelihood (e.g., at 217\,GHz). In temperature, maps are cleaned by template subtraction using higher-frequency maps dominated by thermal dust emission, and by masking bright sources. In polarization, only masking is applied. These steps reduce the residual foreground contamination to a level where explicit modelling in polarization becomes unnecessary and the spectra can be co-added across frequencies to inject in the likelihood a single CMB-only spectrum.

In temperature, residual extragalactic foreground power is still present but can no longer be modelled with specific astrophysical components (such as tSZ, CIB, etc.) because the pre-cleaning steps have coupled the residual emission associated to different contributions. The frequency-dependent residual power is then modelled with a simple power-law form
\begin{equation}
    \mathcal{D}_{\ell,\nu_1,\nu_2}^{\rm fg,TT}(A_{\nu_1,\nu_2}, \gamma_{\nu_1,\nu_2}) = A_{\nu_1,\nu_2} \left( \frac{\ell}{\ell_0} \right)^{\gamma_{\nu_1,\nu_2}},
\end{equation}
with nuisance amplitude parameters $A_{\nu_1,\nu_2}$ and power-law index $\gamma_{\nu_1,\nu_2}$ for each frequency pair. The pivot scale is fixed at $\ell_0 = 1500$. This foreground term is added to the CMB theory before correcting the full model for remaining uncertainties in calibration
\begin{equation}
    \mathcal{D}^{\rm th,,XY}_{\ell,\nu_1,\nu_2}(\theta) = \frac{1}{C^{XY}} \left( \mathcal{D}_{\ell}^{\rm CMB, XY}(\theta_1) + \mathcal{D}_{\ell,\nu_1,\nu_2}^{\rm fg,XY}(\theta_2) \right) \,.
    \label{eq:camspec_th}
\end{equation}
\camspecnpipe does this with three parameters capturing small residual mismatches in gain and polarization efficiency: $C^{TE}$ and $C^{EE}$ allow for rescaling of the TE and EE spectra respectively, and $A_{\rm Planck}$, a global calibration amplitude is applied to all spectra. Like for \plik, this model is compared to the data in a Gaussian likelihood.

We implement all components of the \camspecnpipe\ likelihood in the \mflike\ framework and check that our implementation reproduces the results of R22 as shown in later sections. From this we derive and release the equivalent of \pliklegacy-\lite, \camspecnpipe-\lite, as described in the next section.





\subsection{A foreground-marginalized \camspecnpipe likelihood}
We employ the Gibbs sampling method developed before for ACT, \planck and SPT~\citep{Dunkley2013,ACT:2020frw,dr6ps,Planck_spec_like_2015,Planck_spec_like_2018,2025OJAp....8E..17B,Camphuis:2025uvg} to extract the CMB-only power from the multi-frequency \camspecnpipe dataset. 

For the theory vector of~\autoref{eq:camspec_th}, the Gibbs sampling method builds a Gaussian estimator of the CMB-only bandpowers $\mathcal{D}_{\ell}^{{\rm CMB, XY}}$, marginalized over the secondary parameters $\theta_2$, sampled independently of any cosmological model with parameters $\theta_1$. Gibbs sampling allows us to sample $\mathcal{D}_\ell^{\rm CMB}$ and $\theta_2$ even if we do not have a good method of jointly sampling $(\mathcal{D}_\ell^{\rm CMB},\theta_2)$. In this case, we can easily perform a Gaussian sampling over the CMB bandpowers $\mathcal{D}_\ell^{\rm CMB} | \theta_2$, and we can use Metropolis-Hastings to sample the nuisance parameters $\theta_2 | \mathcal{D}_\ell^{\rm CMB}$.
To do this, we construct the \emph{mapping matrix} $\mathbf{M}$, a rectangular matrix with $M_{ij} = 1$ if the $i$th entry of the CMB vector should contain the bandpowers contained in the $j$th entry of the data vector, and $0$ otherwise. If we have a sample $\mathcal{D}_\ell^{\rm CMB}$, our full model vector becomes
\begin{equation}
    \mathcal{D}_{\ell,\nu_1,\nu_2}(\theta) = \mathbf{M} \mathcal{D}_\ell^{\rm CMB} + \mathcal{D}^{\rm sec}_{\ell,\nu_1,\nu_2}(\theta_2) ,
\end{equation}
where instead of parameterizing $\mathcal{D}_\ell^{\rm CMB}(\theta_1)$ with some Einstein-Boltzmann code based on cosmological parameters $\theta_1$, we now use our Gaussian sample for the CMB bandpowers. Then, our full likelihood becomes
\begin{align}
    -2 \ln \mathcal{L}(\mathcal{D}_\ell^{\rm CMB}, \theta_2)  =  \left( \mathbf{M} \mathcal{D}_\ell^{\rm CMB} + \mathcal{D}^{\rm sec}_{\ell,\nu_1,\nu_2}(\theta_2) - \hat{\mathcal{D}}_{\ell,\nu_1,\nu_2} \right)  \nonumber \\
     \mathbf{\Sigma}^{-1} \left( \mathbf{M} \mathcal{D}_\ell^{\rm CMB} + \mathcal{D}^{\rm sec}_{\ell,\nu_1,\nu_2}(\theta_2) - \hat{\mathcal{D}}_{\ell,\nu_1,\nu_2} \right)^T ,
\end{align}

\noindent with $\hat{\mathcal{D}}_{\ell,\nu_1,\nu_2}$ being our data vector.

For fixed nuisances $\theta_2$, the conditional distribution of the CMB bandpowers, $\mathbb{P}(\mathcal{D}_\ell^{\rm CMB}|\theta_2, \hat{\mathcal{D}}_{\ell,\nu_1,\nu_2})$ is given by
\begin{align}
   -2 \ln \mathbb{P}(\mathcal{D}_\ell^{\rm CMB}|\theta_2, \hat{\mathcal{D}}_{\ell,\nu_1,\nu_2}) = \nonumber\\ 
   \left( \mathcal{D}_\ell^{\rm CMB} - \hat{\mathcal{D}}_\ell^{\rm res} \right)^T & \mathbf{Q}^{-1} \left( \mathcal{D}_\ell^{\rm CMB} - \hat{\mathcal{D}}_\ell^{\rm res} \right) ,
\end{align}

\noindent where $\hat{\mathcal{D}}_\ell^{\rm res}$ is the residual of the data minus the nuisance components, averaged across all cross-spectra, and $\mathbf{Q}$ is the covariance of the CMB bandpowers. We can find these by taking the derivative with respect to the CMB bandpowers, assuming this is a Gaussian distribution with uniform flat priors, giving us
\begin{equation}
    \mathbf{Q} = \mathbf{M}^T \mathbf{\Sigma}^{-1} \mathbf{M} ,
\end{equation}
and
\begin{equation}
\hat{\mathcal{D}}_\ell^{\rm res} = \mathbf{Q}^{-1} \left[ \mathbf{M}^T \mathbf{\Sigma}^{-1} \left( \hat{\mathcal{D}}_{\ell,\nu_1,\nu_2} - \mathcal{D}^{\rm sec}_{\ell,\nu_1,\nu_2}(\theta_2) \right) \right] .
\end{equation}

We can draw a random sample from this Gaussian distribution by taking the Cholesky decomposition of the covariance matrix, $\mathbf{Q} = \mathbf{L} \mathbf{L}^T$, and drawing a random vector
\begin{equation}
    \mathcal{D}_\ell^{\rm CMB} = \hat{\mathcal{D}}_\ell^{\rm res} + \mathbf{L} y
\end{equation}
where $y \sim \mathcal{N}(\mathbf{0},\mathbf{1})$ is a vector drawn from a standard normal distribution. This method means that the vector $\mathcal{D}_\ell^{\rm CMB}$ is drawn from a multivariate normal distribution with a mean and variance imposed by the data vector and covariance matrix.

We can now estimate the nuisance-marginalized CMB bandpowers by Gibbs sampling the conditional distributions $\mathbb{P}(\mathcal{D}_\ell^{\rm CMB}|\theta_2, \hat{\mathcal{D}}_{\ell,\nu_1,\nu_2})$ and $\mathbb{P}(\theta_2 | \mathcal{D}_\ell^{\rm CMB}, \hat{\mathcal{D}}_{\ell,\nu_1,\nu_2})$. Remaining for the nuisance parameters are the six temperature residual foreground parameters, three amplitudes $A_{\nu_1,\nu_2}$ and three power law indices $\gamma_{\nu_1,\nu_2}$, one pair for each of the three TT cross-spectra $143\times143$, $143\times217$, and $217\times217$. We keep the three calibration parameters $A_{\rm planck}$, $C^{TE}$, and $C^{EE}$ fixed to unity in the foreground marginalization, and leave them for the likelihood to measure, as they are fully degenerate with the amplitude of the CMB bandpowers. It is possible to modify the mapping matrix to include any multiplicative calibration parameters and marginalize over them, using a prior to constrain them to avoid the full amplitude degeneracy. We opted however not to do so, as to leave the option for external datasets to jointly marginalize over the \planck dipole calibration measurement. 

\begin{figure}
    \includegraphics[width=\columnwidth]{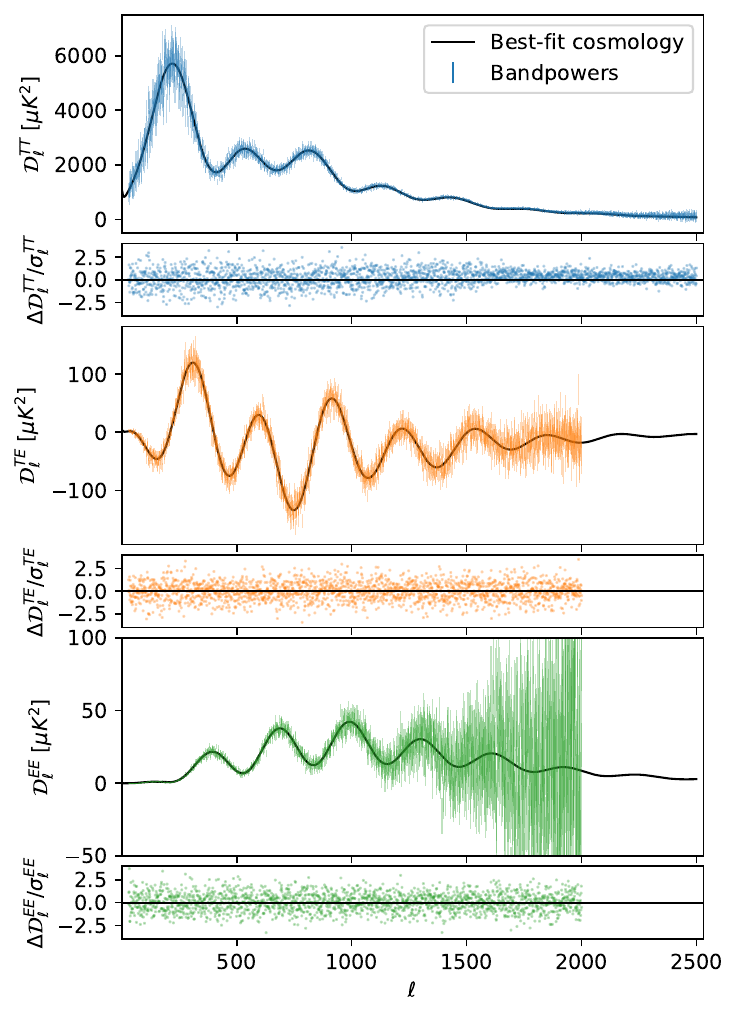}
    \vspace{-0.7cm}
    \caption{CMB foreground-marginalized \camspecnpipe-\lite spectra. The TE and EE spectra appear normal distributed around the best-fitting model, as does the TT spectrum below $\ell < 1500$. At $\ell > 1500$, the TT datapoints are highly correlated and the full behaviour of the data is captured in the covariance matrix.}
    \label{fig:npipe-spectra}
\end{figure}

\begin{figure}
    \includegraphics[width=\columnwidth]{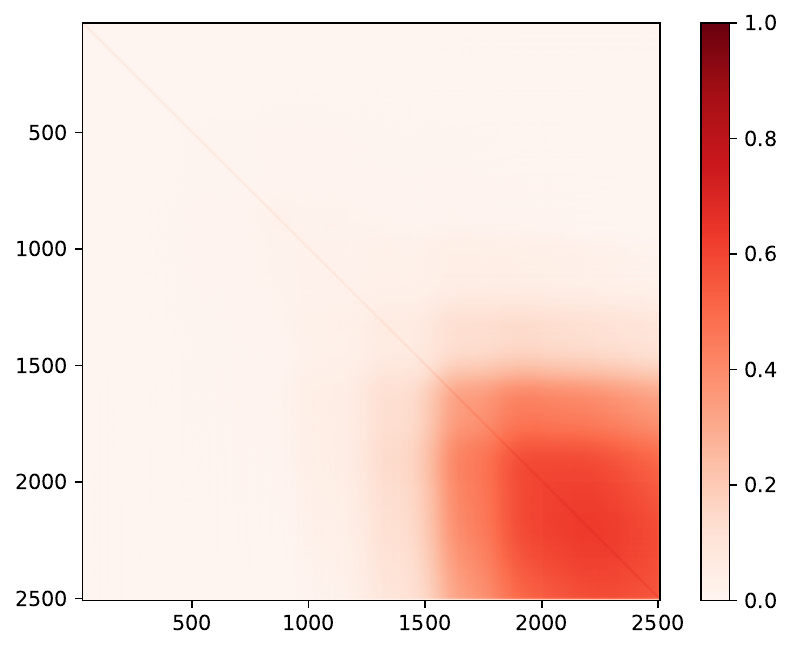}
    \vspace{-0.2cm}
    \caption{The TTTT subblock of the correlation matrix. The marginalization over foregrounds captures a lot of information in the $\ell > 1500$ part of the covariance, leading to high correlations between different bins.}
    \label{fig:correlation-TTTT}
\end{figure}

We show the extracted power spectra from the marginalization procedure in~\autoref{fig:npipe-spectra}. The CMB spectra are shown alongside the best-fitting \lcdm model; a deviation from the best-fit cosmology of the TT bandpowers is visible at $\ell > 1500$ and caused by the foreground-induced correlation between the datapoints which is fully captured in the CMB covariance matrix (similarly to the ACT DR6 small-scale temperature, see Figure 34 of~\citealp{dr6ps}). We show the TTTT block of the correlation matrix in~\autoref{fig:correlation-TTTT}, the remaining blocks are mostly zero or diagonal, as there is no residual TE/EE foregrounds, nor is there much important cross-spectrum covariance.

\begin{figure}
    \includegraphics[width=\linewidth]{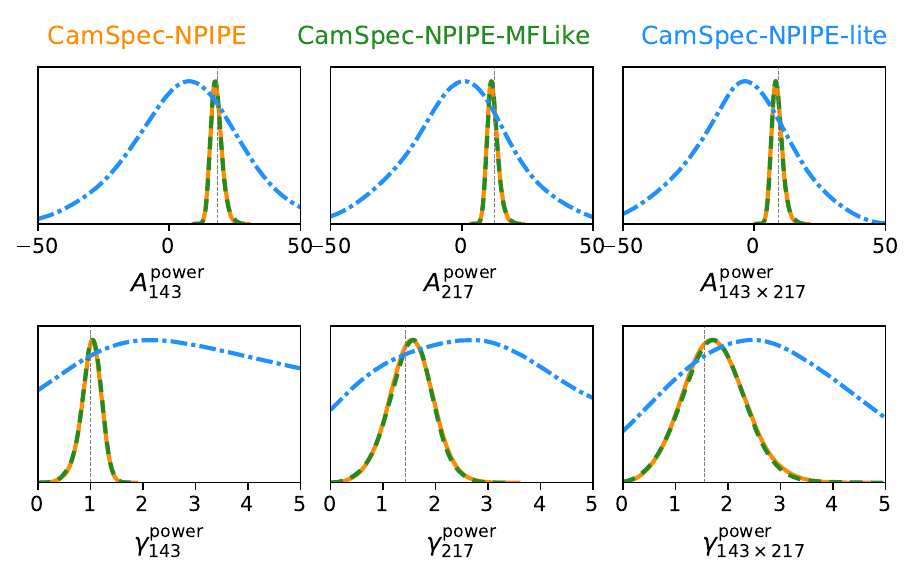}
    \vspace{-0.7cm}
    \caption{Posterior distributions of the foreground nuisance parameters from the multi-frequency likelihood (orange for the original \camspecnpipe likelihood and green for our implementation in \mflike), and from the foreground marginalization procedure (blue). Due to the fact that each foreground appears in only one spectrum each, they are fully degenerate in the foreground marginalization procedure, and thus highly unconstrained. Orange and green posteriors are obtained with a fit to the full \planck TT/TE/EE+lowT+lowE dataset but receive contribution only from the high-$\ell$ \camspecnpipe likelihood. The vertical lines indicate the best-fitting values from the \texttt{CamSpec-NPIPE} chain (original likelihood, shown in orange).}
    \label{fig:fg-params}
\end{figure}

In previous applications, this marginalization procedure relied on the possibility to disentangle (most) frequency-dependent effects from the black-body CMB signal --- i.e., the amplitude of a given foreground component would increase or decrease with frequency while the CMB amplitude would remain constant. In the \camspecnpipe model this is no longer possible because the foreground residual power is distinct across frequencies. We find that the method still works and able to lock the CMB signal across spectra but we lose constraining power on the foreground nuisance parameters with respect to the full likelihood. As shown in \autoref{fig:fg-params} we recover consistent but broader posteriors for the foregrounds (similarly to the Poisson source terms in the case of \pliklegacy, PL20). \autoref{fig:fg-params} also shows that, differently to the full likelihood, for the foreground marginalization procedure we let the amplitude of parameters go negative. We found that opening up the amplitude parameters like this, while not necessarily physical, improved the agreement between the two likelihoods. We explain this by the fact that we model the likelihood as a Gaussian, and without the frequency information the amplitudes will broaden in such a way that imposing a positivity prior truncates one tail of the posterior mode.

After marginalization, we take the resulting TT, TE, and EE samples, and compute the multivariate Gaussian mean $\hat{\mathcal{D}}_{\ell}$ and covariance $\mathbf{\Sigma}$ of these samples. We built a simple Gaussian likelihood, \camspecnpipe-\lite\footnote{\onreview{Available at \url{https://github.com/HTJense/camspec_npipe-lite}}}, as
\begin{align}
    -2 \ln \mathcal{L}(\mathcal{D}_\ell^{\rm CMB}, c^{XY}) = &  \left(\mathcal{D}_\ell^{\rm CMB}/c^{XY} - \hat{\mathcal{D}}_{\ell} \right) \nonumber\\ & \mathbf{\Sigma}^{-1} \left(\mathcal{D}_\ell^{\rm CMB} / c^{XY} - \hat{\mathcal{D}}_{\ell} \right)^T , 
\end{align}
where $c^{XY}$ is a set of calibration parameters for the $XY$ spectra: $c^{TT} = {A_{\rm Planck}}^2$, $c^{TE} = {\rm cal}^{TE} {A_{\rm Planck}}^2$, $c^{EE} = {\rm cal}^{EE} {A_{\rm Planck}}^2$. Thus the entire likelihood only depends on three remaining parameters $\left\{ A_{\rm planck}, {\rm cal}^{TE}, {\rm cal}^{EE} \right\}$ that need to be included in likelihood evaluations. \\

\section{Cosmological constraints} \label{sec:cosmology}


\subsection{Parameter run setup} \label{sec:data-combos}


We implement our likelihoods in \cobaya~\citep{Cobaya2,Cobaya1} and we use it to perform cosmological inference with Monte Carlo Markov Chain runs. We use the Einstein-Boltzmann solver code \camb for generating the lensed CMB power spectrum with the accuracy settings described in~\citet{dr6ext}, even when only using \planck data, for consistency. We run our chains to a convergence criterion of $R-1 < 0.01$, discarding 30\% of our samples as a burn-in fraction. 

As a baseline comparison, we derive constraints on the \lcdm model parameters from the official \pliklegacy and \camspecnpipe likelihoods (as implemented in~\cobaya), from our \mflike implementations, as well as our \camspecnpipe-\lite likelihood. We vary the standard set of six base \lcdm parameters: the baryon and cold dark matter densities, $\Omega_b h^2$ and $\Omega_c h^2$, the amplitude and spectral index of primordial curvature scalar perturbations, $A_s$ and $n_s$ defined at a pivot scale $k_0 = 0.05~\rm{Mpc}^{-1}$, an approximation of the angular scale at recombination, $\theta_{\rm MC}$, and the optical depth to reionization, $\tau_{\rm reio}$. We assume flatness and the existence of massive neutrinos with a total fixed mass of $0.06~\rm{eV}$. We also test select extension models, including variations in the lensing peak smearing $\Lambda$CDM$+A_L$ (with $A_L=1$ in \lcdm;~\citealp{Calabrese2008}), in the effective number of relativistic species $\Lambda$CDM$+N_{\rm eff}$ ($N_{\rm eff}=3.044$ in \lcdm;~\citealp{2021JCAP...04..073B,2024JCAP...06..032D}), and in the running of the spectral index of primordial scalar curvature perturbations $\Lambda$CDM+$\mathrm{d} n_s/\mathrm{d} \ln k$ ($\mathrm{d} n_s/\mathrm{d} \ln k=0$ in \lcdm;~\citealp{1995PhRvD..52.1739K}). 

We follow the naming convention for data combinations used by PL20 and \citet{dr6ps}.

\begin{itemize}[topsep=1pt,leftmargin=12pt]
\item[(i)] We refer to \planck TT/TE/EE when using the data combination of TT, TE, and EE in the multipole range ${30<\ell< 2508}$ from either \pliklegacy or \camspecnpipe likelihoods. To these we always add the large-scale temperature data of the \texttt{Commander} likelihood from PL20 at $\ell<30$ and denote it with `lowT'. We further add the EE large-scale polarization data from the \planck \texttt{Sroll2} likelihood~\citep{sroll2} with `Sroll2'; we sometimes replace this full likelihood with an equivalent Gaussian prior $\tau_{\rm reio} = (5.44 \pm 0.73) \times 10^{-2}$ for quicker evaluations and label this `lowE'.\footnote{As shown in other ACT and SPT works, this prior is enough to break the $A_s$-$\tau_{\rm reio}$ degeneracy.} 
\item[(ii)] In some cases we investigate the constraints from the \planck largest scales only --- the so-called `cut' data ranges, with the high-$\ell$ multipole range chosen by~\cite{dr6ps} of $\ell<1000/600/600$ in TT/TE/EE. We explicitly call these truncated likelihoods \pliklegacy-\lite-\cut and \camspecnpipe-\lite-\cut, referring to the full ranges otherwise. 
\item[(iii)] We combine the two \planck \lite-\cut likelihoods with the small-scale data from ACT DR6. In these cases, we use the nomenclature from L25, calling \pact the combination of either \planck \lite-\cut likelihood with the ACT DR6-\lite likelihood, as well as the large-scale temperature and polarization measurements from \texttt{Commander} and \texttt{Sroll2}. We explicitly differentiate between \pact (\pliklegacy) and \pact (\camspecnpipe).
\end{itemize}


\subsection{Parameter results} \label{subsec:parameter-results}

\subsubsection{\camspecnpipe-\lite vs full multi-frequency}

\begin{figure}
    \includegraphics[width=\linewidth]{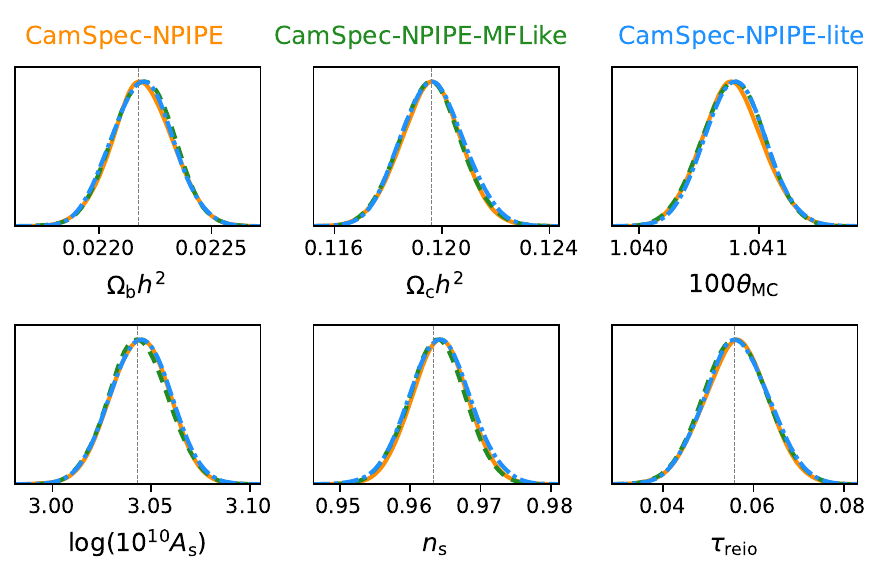}
    \vspace{-0.5cm}
    \caption{Posteriors of the \lcdm cosmological parameters recovered from different implementations of the \camspecnpipe likelihood. We find very good agreement in all five free parameters (we infer the optical depth $\tau_{\rm reio}$ through a prior), and see only a slight widening of $9\%$ on the constraint on $n_s$ from the \lite likelihood due to the foreground marginalization. All posteriors are obtained with the full \planck TT/TE/EE+lowT+lowE dataset. The vertical lines indicate the best-fitting values from the \texttt{CamSpec-NPIPE} chain (original likelihood, shown in orange).}
    \label{fig:cosmology}
\end{figure}

To validate our implementation of the \camspec likelihood, we perform cosmological parameter inference with two versions of the full multi-frequency \camspecnpipe likelihood: the original version from R22 and our \mflike implementation, finding perfect agreement both at parameter posterior level and $\chi^2$ evaluation. We also compare the results from these full likelihoods to those derived from our compressed \camspecnpipe-\lite likelihood. The constraints on cosmological parameters are shown in~\autoref{fig:cosmology}. We show an excellent recovery of all five cosmological parameters\footnote{The optical depth $\tau_{\rm reio}$ is measured in all cases by \texttt{Sroll2}.}, with only a minor ($\lesssim 10\%$) widening of the distribution of the primordial power spectrum tilt $n_s$ from the \lite likelihood.

\begin{figure}
    \centering
    \includegraphics[width=\columnwidth]{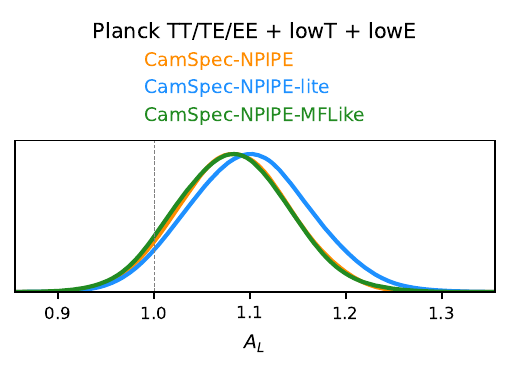}
    \includegraphics[width=\columnwidth]{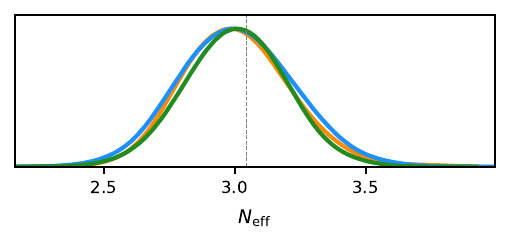}
    \includegraphics[width=\columnwidth]{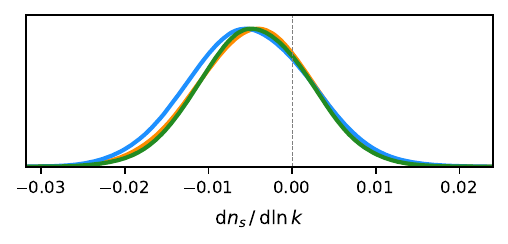}
    \vspace{-0.5cm}
    \caption{Comparison of the constraints on select single-parameter extensions to \lcdm as obtained from the full \planck TT/TE/EE+lowT+lowE dataset using the original \camspecnpipe likelihood (orange), our implementation in \mflike (green), and our \camspecnpipe-\lite likelihood (blue). We show the constraints on the lensing peak smearing amplitude $A_L$ (top panel), the effective number of relativistic species $N_{\rm eff}$ (middle panel), and the running of the spectral index ${\rm d} n_s / {\rm d} \ln k$ (bottom panel). We observe that foreground marginalization has little effect on the constraints of extension models. The constraints from the \lite likelihood are consistent and at most $5-10\%$ wider than those from the full likelihoods. The vertical lines show the expected values of these parameters in \lcdm.}
    \label{fig:extensions}
\end{figure}

The constraints on the parameters of the three extended models tested here are compared between the full and \lite likelihoods in~\autoref{fig:extensions}, again finding good agreement between the different results. We recover the exact same posteriors between our \camspecnpipe-\mflike and the official \camspecnpipe implementations, and good agreement between those and the parameter posteriors obtained when using \camspecnpipe-\lite. Also in this case we observe a minor broadening of parameter uncertainties ($5-10\%$) from the \lite likelihood due to the foreground marginalization step. For $A_L$ we note that the broadening is more enhanced on the high-end tail of the posterior, the full 2-dimensional posterior shows that this is down to a one-sided broadening of parameters along the ${A_L-\Omega_b h^2}$ degeneracy line, instead of a shift in central value.

\subsubsection{Comparison with \pliklegacy}

\begin{figure}
    \includegraphics[width=\linewidth]{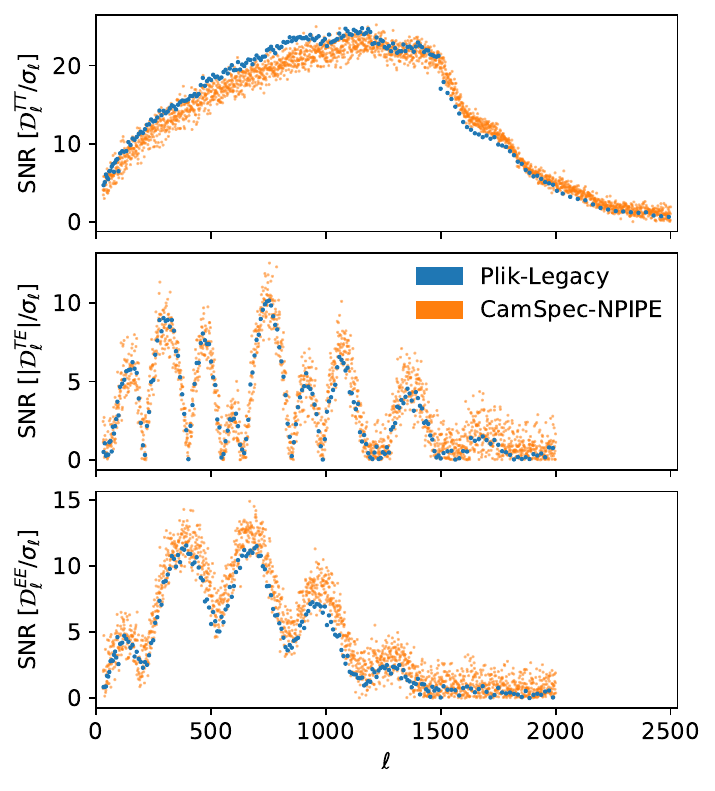}
    \vspace{-0.7cm}
    \caption{Comparison of the signal-to-noise ratios $\mathcal{D}_\ell / \sigma_\ell$ of \pliklegacy (blue) and \camspecnpipe (orange) after foreground-marginalization. The \plik data points contain an additional factor of $\sqrt{\Delta \ell}$ to correct for the bin widths of the data, while the \npipe is plotted with an additional $1\%$ error margin since the likelihood marginalizes over additional TE and EE calibration factors. \onreview{The constraining power in $TT$ from \pliklegacy is higher at $\ell < 1200$ due to the inclusion of the $100 \times 100$ spectrum, which is absent in \camspecnpipe.}}
    \label{fig:plik-npipe-comparison}
\end{figure}

\begin{figure*}
    \includegraphics[width=0.9\linewidth]{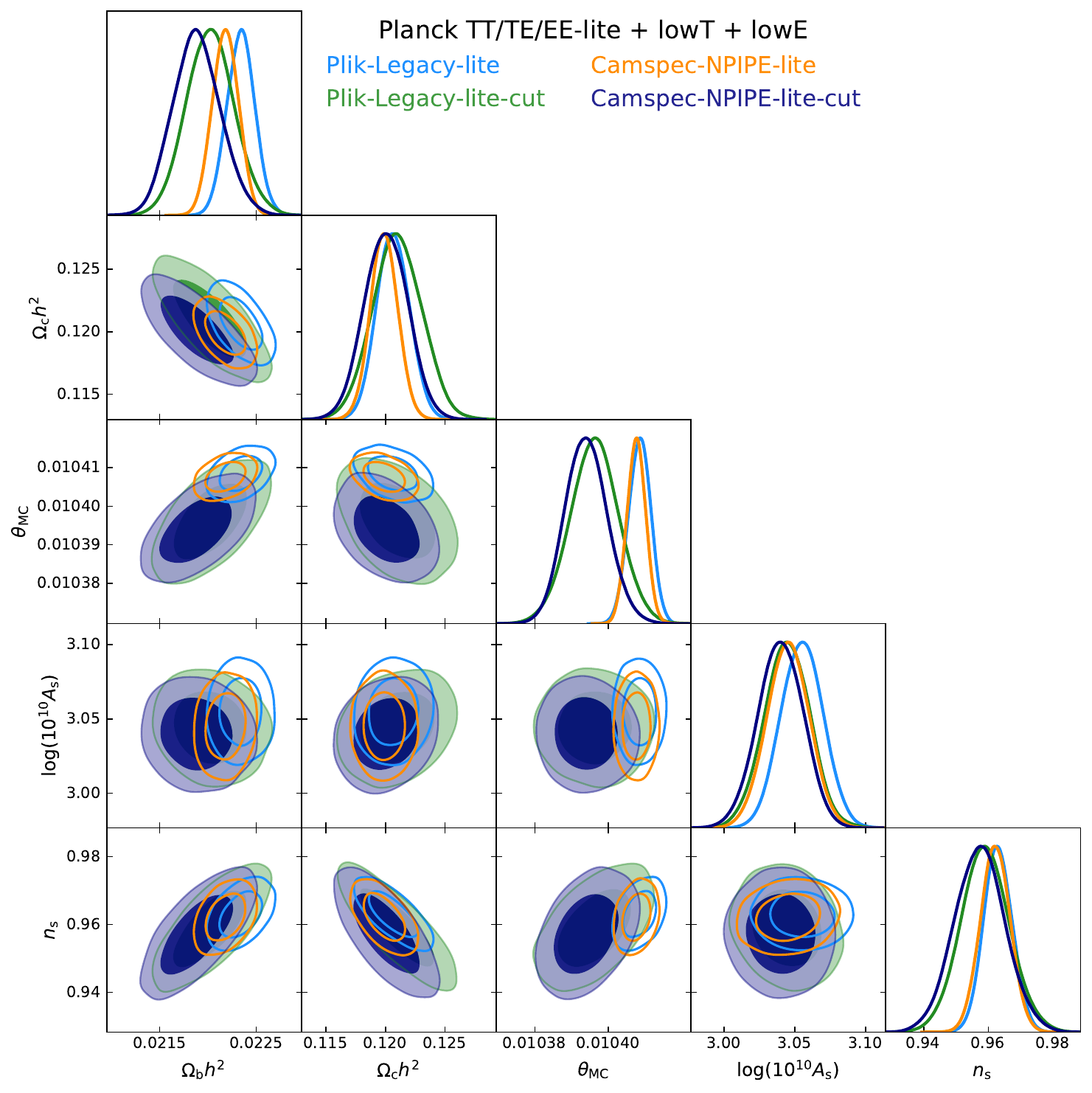}
    \caption{Comparison of \lcdm parameters as obtained from \pliklegacy and \camspecnpipe likelihoods, using the full dataset (blue and orange respectively) and a restricted range keeping only the large scales with $\ell<1000/600/600$ in TT/TE/EE (green and navy respectively). The results from the full datasets (open contours) do not overlap as much in parameter space as the cut likelihoods (filled contours). \onreview{The numerical values of these constraints are given in~\autoref{tab:full-cut-parameter-table}.}\\}
    \label{fig:npipe-comparison-cut}
\end{figure*}

Differences between \pliklegacy and \camspecnpipe cosmologies are expected due to small differences in the amount of data used and many differences in data characterization; these have been discussed at length in PL20, EG21 and R22. To summarize, parameters in \lcdm are consistent but with the mean of the posteriors shifting by fractions of $\sigma$s and in the case of $\Omega_b h^2$ by $1\sigma$ (see~\autoref{fig:npipe-comparison-cut}). Following R22, to give a quantitative comparison we confront either best-fitting cosmology to the either dataset. This is shown in~\autoref{tab:chi2-best-fit-crosscomparison}. We find a $\Delta \chi^2 = 2.88$ between the \camspecnpipe-\lite and \pliklegacy-\lite best-fit \lcdm theories when compared to the \camspecnpipe-\lite data, and $\Delta \chi^2 = 3.83$ when compared to the \pliklegacy-\lite data, showing good agreement between the two cosmological solutions. For these calculations we keep the cosmological parameters fixed to the best-fit values but minimize for calibrations.


\begin{table*}
    \centering

    \begin{tabular}{c l l l l l l}
        \toprule
        & \multicolumn{6}{c}{Theory} \\
        & \multicolumn{2}{c}{\planck TT/TE/EE+lowT+lowE} & \multicolumn{2}{c}{\planck-\cut TT/TE/EE+lowT+lowE} & \multicolumn{2}{c}{\pact} \\
        & PL-\lite & CN-\lite & PL-\lite-\cut & CN-\lite-\cut & PL-\lite-\cut & CN-\lite-\cut \\
        \cmidrule{2-7}
        High-$\ell$ dataset & & & & &\\    
        \cmidrule{2-3}
        PL-\lite & 583.27 & 587.10 (3.83) \\
        CN-\lite & 6745.63 (2.88) & 6742.75 \\
        \cmidrule{2-3}
        \cmidrule{4-5}
        PL-\lite-\cut & & & 216.41 & 218.17 (1.76) \\
       CN-\lite-\cut & & & 2332.46 (1.75) & 2330.71 \\
        \cmidrule{4-5}
        \cmidrule{6-7}
        \pact (PL-\lite-\cut) & & & & & 378.73 & 379.47 (0.74) \\
        \pact (CN-\lite-\cut) & & & & & 2499.18 (0.39) & 2498.79  \\
        \bottomrule
    \end{tabular}
    
    \caption{A comparison of best-fit \lcdm solutions obtained with different \planck likelihoods fitted to either \pliklegacy or \camspecnpipe datasets (shortened here into PL and CN, respectively). The rows give the $\chi^2$ value of a given high-$\ell$ dataset for the best-fitting cosmology from each column, with the number in brackets referring to the $\Delta\chi^2$ with respect to the self best-fitting curve of each dataset. There is good agreement between the two \planck products and remaining differences decrease for a cut multipole range or when combined with other CMB data. Similar to R22, to calculate the $\chi^2$ we minimize for the calibration parameters, while keeping the cosmological parameters fixed.\\}
    \label{tab:chi2-best-fit-crosscomparison}
    \end{table*}

The small increase in data going from the Legacy maps to NPIPE and the pre-cleaning steps allowing to retain more sky fraction in \camspecnpipe translates into tighter limits for both \lcdm and extended model parameters. To localize where in spectrum space the additional constraining power is coming from, we show a comparison of the \pliklegacy-\lite and \camspecnpipe-\lite per-$\ell$ constraining power in \autoref{fig:plik-npipe-comparison}, calculating the signal-to-noise ratio per bin in the spectra, corrected for the different bin sizes and the slight difference in the treatment of polarization efficiencies. We see that the \camspecnpipe-\lite data has better constraining power in TE and EE, as well as in TT at $\ell > 1500$. Better polarization and improved small scales in temperature are particularly important for measuring $n_s$ and extended model parameters.

\subsubsection{Parameters from restricted multipole range}

We now test a restricted multipole range, cutting the \planck power spectra in TT/TE/EE at $\ell < 1000/600/600$, which is the point at which ground-based experiments currently probe power spectra to higher precision than \planck. These criteria were first set out in \citet{dr6ps} for ACT DR6 and later re-used in \citet{Camphuis:2025uvg} for the combination of \planck, ACT and SPT-3G.

We show the constraints on \lcdm parameters in \autoref{fig:npipe-comparison-cut} \onreview{and their numerical values in~\autoref{tab:full-cut-parameter-table}}, comparing the \camspecnpipe and \camspecnpipe-\cut constraints versus the equivalent for \plik. The restricted multipole range gives very similar parameters between the two likelihoods, improving on the agreement from the full dataset. The largest shift is of $\sim 0.4\sigma$ in $\Omega_b h^2$ and the $\Delta \chi^2$ with respect to the fit to the other dataset is reduced by $\sim 1-2$ points (see Table 2).

\begin{table*}
\centering
    \begin{tabular}{l c c|c c}
        \toprule
        & \textbf{\pliklegacy-\lite} & \textbf{\camspecnpipe-\lite} & \textbf{\pliklegacy-\lite-\cut} & \textbf{\camspecnpipe-\lite-\cut} \\
        \midrule
        $\Omega_b h^2$ & $(2.237 \pm 0.015) \times 10^{-2}$ & $(2.218 \pm 0.014) \times 10^{-2}$ & $(2.203 \pm 0.027) \times 10^{-2}$ & $(2.188 \pm 0.025) \times 10^{-2}$ \\
        $\Omega_c h^2$ & $(12.01 \pm 0.14) \times 10^{-2}$ & $(11.99 \pm 0.12) \times 10^{-2}$ & $(12.07 \pm 0.22) \times 10^{-2}$ & $(12.00 \pm 0.18) \times 10^{-2}$ \\
        $\theta_\mathrm{MC}$ & $(104.087 \pm 0.031) \times 10^{-4}$ & $(104.077 \pm 0.026) \times 10^{-4}$ & $(103.952 \pm 0.070) \times 10^{-4}$ & $(103.944 \pm 0.057) \times 10^{-4}$ \\
        $n_s$ & $0.9651 \pm 0.0043$ & $0.9622 \pm 0.0046$ & $0.9594 \pm 0.0078$ & $0.9575 \pm 0.0077$ \\
        $\log(10^{10} A_s)$ & $3.054 \pm 0.014$ & $3.046 \pm 0.016$ & $3.043 \pm 0.016$ & $3.040 \pm 0.016$ \\
        \bottomrule
    \end{tabular}
    \caption{\onreview{Constraints on the \lcdm basic varying parameters from \pliklegacy-\lite and \camspecnpipe-\lite, comparing the full (left) and restricted multipole (right) ranges. These numerical values correspond to the results shown in~\autoref{fig:npipe-comparison-cut}. \\}}
    \label{tab:full-cut-parameter-table}
\end{table*}

\section{Combining Planck with other CMB datasets} \label{sec:p-act-cosmology}

One motivation for this work is to generate a \camspecnpipe product that can be easily and correctly combined at likelihood level with other CMB datasets. The foreground marginalization performed here allows to properly treat the foreground contribution in the large-scale \npipe data when combining the cut likelihood with a different small-scale measurement, for example from ACT DR6. 

With this product now available, we derive here \pact cosmological constraints for different \planck likelihoods in combination with ACT DR6 as defined in \autoref{sec:data-combos}, to check if the specific choice of \planck is important in these combined results. Throughout this section, we only combine \lite likelihoods.


\subsection{Constraints on \lcdm}

\begin{figure*}
    \centering
    \includegraphics[width=0.9\linewidth]{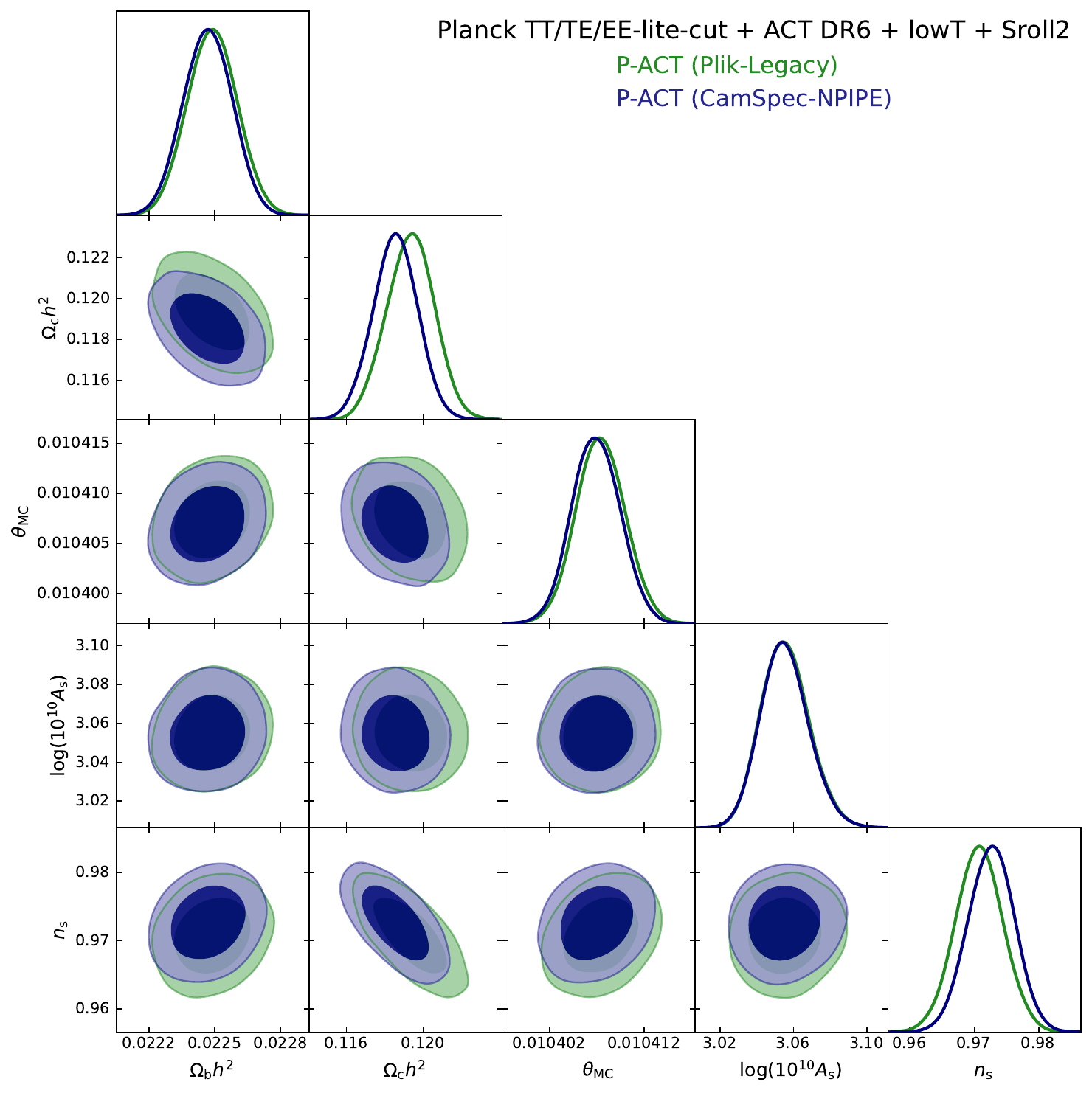}
    \caption{\lcdm constraints from the \pact combination with either \pliklegacy (green) or \camspecnpipe (dark blue) in all five cosmological parameters. We find excellent agreement in between either choice in the \pact combination. Compared to~\autoref{fig:npipe-comparison-cut}, we notice that there is less difference in most parameters.\\}
    \label{fig:pact-pr3-pr4-comparison}
\end{figure*}

We show a comparison between the constraints obtained with \pliklegacy and \camspecnpipe in combination with small-scale ACT DR6 data, in \autoref{fig:pact-pr3-pr4-comparison}.

Overall, we see that the agreement between \pliklegacy and \camspecnpipe is maximum when combined with ACT DR6, reducing their difference to $\Delta \chi^2 = 0.4-0.7$ in the \pact combination. This is partially due to having cut the \planck likelihoods and partially to the additional constraining power from ACT. 
The $1.0 \sigma$ difference in $\Omega_b h^2$ between the two full \planck datasets goes down to $0.4\sigma$ for the \planck cut cases and then $0.15 \sigma$ when combining with ACT DR6. In the \pact combination, the largest parameter difference is a $0.7 \sigma$ difference in $\Omega_c h^2$. We compare the numerical values of the \lcdm parameter constraints in~\autoref{tab:lcdm-parameter-table}.

\begin{table*}
\centering
    \begin{tabular}{l c c}
        \toprule
        & \textbf{P-ACT (\pliklegacy)} & \textbf{P-ACT (\camspecnpipe)} \\
        \midrule
        $\Omega_b h^2$ & $(2.249 \pm 0.011) \times 10^{-2}$ & $(2.247 \pm 0.011) \times 10^{-2}$ \\
        $\Omega_c h^2$ & $(11.93 \pm 0.12) \times 10^{-2}$ & $(11.85 \pm 0.12) \times 10^{-2}$ \\
        $\theta_\mathrm{MC}$ & $(104.074 \pm 0.026) \times 10^{-4}$ & $(104.069 \pm 0.025) \times 10^{-4}$ \\
        $n_s$ & $0.9708 \pm 0.0037$ & $0.9726 \pm 0.0036$ \\
        $\log(10^{10} A_s)$ & $3.055 ^{+0.012}_{-0.014}$ & $3.055 ^{+0.012}_{-0.014}$ \\
        \midrule
        $H_0$ & $67.62 \pm 0.51$ & $67.86 \pm 0.48$ \\
        $\sigma_8$ & $0.8146 \pm 0.0063$ & $0.8125 \pm 0.0062$ \\
        $\Omega_m$ & $(31.03 \pm 0.72) \times 10^{-2}$ & $(30.63 \pm 0.67) \times 10^{-2}$ \\
        $A_s e^{-2 \tau}$ & $(1.883 \pm 0.012) \times 10^{-9}$ & $(1.879 \pm 0.013) \times 10^{-9}$ \\
        \bottomrule
    \end{tabular}
    \caption{Constraints on the \lcdm basic varying (top) and key derived (bottom) parameters from \pact, comparing using \pliklegacy\lite and \camspecnpipe\lite in combination with ACT.\\}
    \label{tab:lcdm-parameter-table}
\end{table*}

\subsection{Constraints on extension models}

\begin{figure}
    \centering
    \includegraphics[width=\columnwidth]{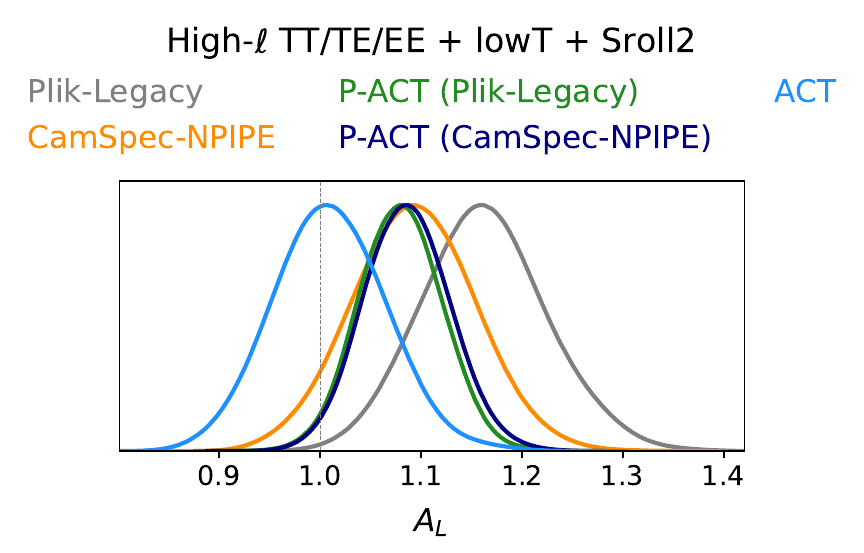} \\
    \includegraphics[width=0.76\columnwidth]{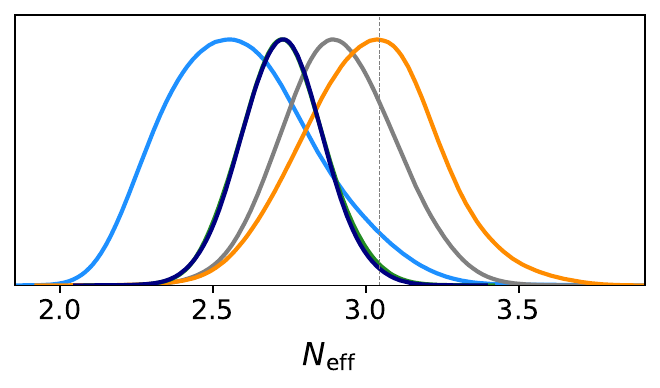}
    \caption{Comparisons of the ACT, \planck, and \pact constraints on $A_L$ and $N_{\rm eff}$ for the \pliklegacy and \camspecnpipe likelihoods. In both cases, the \pact combination gives identical results for either \planck likelihood.}
    \label{fig:pact-extension-comparison}
\end{figure}

We now compare the constraints on the extended models $\Lambda$CDM$+A_L$ and $\Lambda$CDM$+N_{\rm eff}$. 
We expect that the $A_L$ extension carries significant weight from the \planck measurement of large angular scales, while the $N_{\rm eff}$ extension should be dominated by the ACT measurement of small scales. As such, these two extensions are good proxies for the effects of combining the two different datasets. The marginal constraints on the extension parameters for the different \planck likelihoods and their combinations with ACT are shown in~\autoref{fig:pact-extension-comparison}.

We see that the combination of \planck with ACT removes completely the effect of the choice of \planck likelihood on the constraint of the extension parameter. Both $A_L$ and $N_{\rm eff}$ are recovered to the same value, with minimal differences in other parameters.

We extend this test to run the CMB-SPA combination of~\cite{Camphuis:2025uvg} and we find the same conclusions, the results are insensitive to the choice of \planck.

\subsection{Evidence for evolving dark energy}

The recent DESI DR2 results have generated a lot of interest in the prospects of dynamical dark energy, with a tentative statistical preference for $w_0w_a$ over \lcdm~\citep{DESI-DR2,DESI1_DR2} where $w_0$ and $w_a$ parametrize the present-day value of the dark energy equation of state and its variation with time, respectively. In \lcdm $w_0=-1, w_a=0$~\citep{2001IJMPD..10..213C,2003PhRvL..90i1301L}. In light of the vast range of dataset combinations that have been explored for this model, and their difference in statistical preference (see e.g.,~\citealp{DESI_ACT_w0wa}), we test here both the use of \camspecnpipe instead of \pliklegacy and the effects of foreground marginalization for the \camspecnpipe in the combination \texttt{P-ACT-LBS}\footnote{\texttt{P-ACT-LBS} is defined as in \cite{dr6ext} as the combination of large-scale CMB measurements from \planck \texttt{Commander} for TT, \texttt{Sroll2} for EE, \planck for the intermediate scales ${30 < \ell < 1000/600/600}$ in TT/TE/EE, ACT DR6 for the range ${600 < \ell < 6500}$, the \planck NPIPE + ACT DR6 lensing measurement, BAO measurements from DESI DR2, and supernovae from the Pantheon+ sample.} on the $w_0w_a$ constraints and statistical preference over \lcdm. In deriving best-fit points in likelihood space we use the \texttt{BOBYQA} minimizer~\citep{Bobyqa1,Bobyqa2,Bobyqa3}.

\begin{figure}[htp]
    \centering
    \includegraphics[width=\columnwidth]{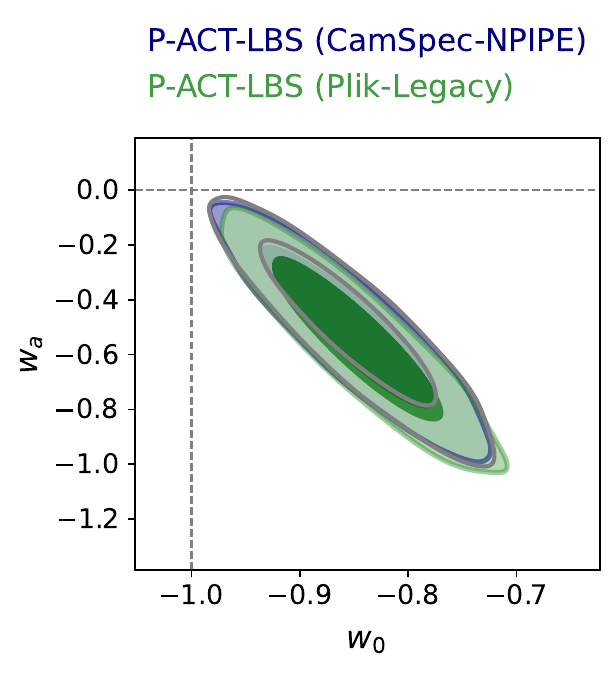}
    \caption{Constraints on $w_0w_a$ from \texttt{P-ACT-LBS} using either \pliklegacy (green) or \camspecnpipe (dark blue). The dashed lines denote \lcdm at $w_0 = -1, w_a = 0$. The empty grey contour uses the same data combination but with non-marginalized \camspecnpipe likelihood. The contours are insensitive to the choice of \planck.}
    \label{fig:pact-w0wa-comparison}
\end{figure}

After foreground marginalization, we use Wilks' theorem to find that the statistical preference for $w_0 w_a$ is $2.5 \sigma$ when using \pliklegacy-\lite-\cut, or $2.4 \sigma$ when using the foreground-marginalized \camspecnpipe-\lite-\cut data. 
With a simple truncation of the full multi-frequency likelihood --- leaving the foreground unconstrained --- for \camspecnpipe-\cut we find $2.3 \sigma$. We conclude that the P-ACT evidence for $w_0 w_a$ does not depend on the choice of \planck (see also~\autoref{fig:pact-w0wa-comparison}) and that the treatment of the nuisance parameters changes this number by $\sim 0.1\sigma$. We note that \cite{DESI_ACT_w0wa} found the same preference when using the \pact combination with \pliklegacy; their combination with \camspecnpipe gives a $2.8\sigma$ preference for $w_0 w_a$ but assumes a substantially different dataset in both $\ell$ cuts ($\ell<2000/1000/1000$ in TT/TE/EE) and the choice of low-$\ell$ EE likelihood and therefore it is not directly comparable with the \pact results obtained with \pliklegacy. 

We summarize these numbers in~\autoref{tab:w0wa-preference} and note that they fluctuate at the level of $0.2-0.3\sigma$ depending on choices of low-$\ell$ EE likelihood, minimizer used for deriving the best-fit points and accuracy settings in theory calculations. The threshold of $3\sigma$ used to claim tentative evidence is therefore conditioned to several assumptions made in the analysis, including the treatment of foregrounds.

\begin{table}[t!]
    \centering
    \begin{tabular}{c c c}
        \toprule
        \texttt{P-ACT-LBS} for \planck likelihood & $\Delta \chi^2$ & Preference \\
        \midrule
        \pliklegacy-\lite-\cut & $-8.8$ & $2.5 \sigma$ \\
        \camspecnpipe-\lite-\cut & $-8.1$ & $2.4 \sigma$ \\
        \camspecnpipe-\cut & $-7.9$ & $2.3 \sigma$ \\
        \bottomrule
    \end{tabular}
    \caption{Summary of the statistics for $w_0 w_a$ over \protect\lcdm from the \texttt{P-ACT-LBS} combination using different \protect\planck likelihoods. The $\Delta \chi^2$ column lists the $\chi^2_{w_0w_a} - \chi^2_{\rm \Lambda CDM}$ difference at the Maximum A Posteriori point, and the preference represents the frequentist significance from this $\chi^2$ difference.}
    \label{tab:w0wa-preference}
\end{table}

\section{Conclusions}
\label{sec:conclusions}

We have summarized differences in cosmological results resulting from different combinations of \planck sky maps and likelihood pipelines. To ease the comparison and to build a product that can be correctly combined with other CMB datasets at likelihood level, we have derived a foreground-marginalized \lite version of the \camspecnpipe likelihood. 
We use this new \lite likelihood to localize in CMB spectrum space the additional constraining power present in \camspecnpipe compared to \pliklegacy and then to perform cosmological analysis with a restricted multipole range as now commonly done by ground-based experiments. We show that limits on cosmological parameters, in particular in extended models, are completely insensitive to the choice of \planck maps/likelihood when keeping the \planck large scales in combination with small scales from ACT DR6.\\

\newpage
\section*{Acknowledgements}

We are grateful to ACT collaborators for useful discussions. We thank Ian Harrison and Serena Giardiello for useful feedback. HTJ, MV and EC acknowledge support from the Horizon 2020 ERC Starting Grant (Grant agreement No 849169). JCH acknowledges support from NSF grant AST-2108536, the Sloan Foundation, and the Simons Foundation.




\bibliographystyle{act_titles}
\bibliography{main} 


\label{lastpage}
\end{document}